\documentclass[11pt]{article}
\usepackage{geometry}                
\usepackage{graphicx}
\usepackage{amssymb}
\usepackage{epstopdf}
\def\authorlist#1#2{
    \vskip 0.4in
\begin{center}\begin{large} {\bf  #1 } \end{large}
    \vskip 0.2in
              #2
     \vskip 0.2in
   \end{center}
}

\begin{document}

\title{\textbf{Snowmass 2021 \\ Underground Facilities \& Infrastructure \\ Topical Report on \\ Synergies in Research at Underground Facilities}}

\maketitle

\authorlist{C.~Curceanu, D.~Elsworth, J.~Formaggio, J.~Harms, D.~Robertson, W.~Roggenthen, H.~Wang}{Additional contributors from the community: S. de Jong \\ Document compiled by J.L. Orrell}

\tableofcontents





\section{Introduction}

A broad range of scientific and engineering research is possible in underground laboratories, beyond the physics-focused activities described in the other Underground Facilities and Infrastructure Topical Reports. These areas of research include nuclear astrophysics, geology, geoengineering, gravitational wave detection, biology, and perhaps soon quantum information science. This UF Topical Report will survey those other scientific and engineering research activities that share interest in research-orientated Underground Facilities and Infrastructure. In most cases the breadth and depth of research aims is too large to cover in completeness and references to surveys or key documents for those fields are provided after introductory summaries. Additional attention is then given to shared, similar, and unique needs of each research area with respect to the broader underground research community's Underground Facilities and Infrastructure needs. Where potential conflicts of usage type, site, or duration might arise, these are identified.

\section{Quantum Information Science}
\label{sec:qisynergy}
Writer: Joe Formaggio (MIT)

\subsection{Science goals}
As both nuclear physics and particle physics involve the quantum interactions of many sub-atomic particles, there has always existed a strong interplay between these fields and the study of quantum physics and quantum information systems (QIS).  This interplay has accelerated in recent years, particularly with the emergence of new, highly sensitive technologies, nascent access to quantum computing environments at the {\cal O}(10)-{\cal O}(100)-bit scale, and the use of coherence and entanglement to enhance sensitivity to novel and exotic phenomena.  One unusual area of interplay between the two disciplines that has recently emerged is the role of background radiation and background mitigation on highly sensitive systems such as qubits.

In order to make quantum computing a viable and usable technology, the underlying units of computation --qubits-- must exhibit high fidelity and long coherence times.  Over the past two decades, advances in device design, fabrication, and materials have increased coherence times by almost six orders of magnitude~\cite{UF5-doi:10.1146/annurev-conmatphys-031119-050605}.  Nonetheless, to realize the full promise of quantum computing, far longer coherence times will be needed to achieve the operational fidelity required for fault-tolerant computation. Coherence for superconducting qubits can be spoiled by an excess density of quasi-particles, and quasi-particle densities far exceed what is naively expected from thermal equilibrium.  Recent measurements made by several groups~\cite{UF5-nature584,UF5-McEwen:2021wdg,UF5-Wilen:2020lgg,UF5-Cardani:2020vvp} have shown that one creeping contribution to quasi-particle poisoning appears due to ionizing radiation stemming from external gamma radiation, cosmic rays, and radiogenic contamination of materials surrounding the qubit.  This source of quasi-particle poisoning is particularly worrisome for QIS applications, since ionizing radiation appears to affect multiple qubits simultaneously~\cite{UF5-McEwen:2021wdg,UF5-Wilen:2020lgg}.  Since quantum error correction (QEC) schemes --necessary for scaling quantum computing-- rely on qubits to exhibit random, uncorrelated errors, correlated error sources would render such schemes ineffective.  Ionizing radiation has been deemed "catastrophic" because of its ability to potentially circumvent traditional QEC algorithms.

\subsection{Current underground-based research}
Fortunately, a number of techniques are being devised to reduce the impact of radiation.  These include spatially distributed error correction schemes~\cite{UF5-Xu:2022wcg}, material engineering to promote phonon down-conversion~\cite{UF5-Catelani:2021ygo,UF5-Iaia:2022jsh}, and developing a more comprehensive understanding of the underlying microscopic physics that leads to quasi-particle poisoning~\cite{UF5-Liu:2022aez,UF5-Pan:2022kpz}.  As cosmic ray radiation is a non-negligible portion of the environmental radiation, underground facilities may also provide a unique resource for studying quasi-particle poisoning in radiation-quiet environments.  Here ``underground facilities" is a catchall for the decades of experience in the nuclear and particle physics communities for background mitigation and suppression.  Underground laboratories specifically may offer perhaps one of the only ways to effectively reduce the cosmic ray flux impingent upon a multi-qubit system.  Although precise models of how cosmic rays impact multi-qubit systems are still being studied, preliminary measurements on similar systems have shown that significant overburden could improve qubit performance~\cite{UF5-Gusenkova:2022jvl,UF5-Cardani:2020vvp}.

\subsection{Underground usage and needs}
Underground facilities could be excellent locations for studying qubit systems in low background environments, with relatively modest investments in infrastructure and resources.  Small facilities have already started to emerge in both deep- and shallow sites, such as FERMILAB, Pacific Northwest National Laboratory, and Gran Sasso.  As qubit systems operate at cryogenic temperatures, such facilities would need to sustain use of dilution refrigerators underground.  Such systems are already in place for several dark matter experiments that operate at cryogenic temperatures (e.g. EDELWEISS~\cite{UF5-EDELWEISS:2020fxc}).

\subsection{Overlap with the high energy physics community}
The quantum information science (QIS) community pursuing superconducting qubits for quantum computing, has meaningful overlap with quantum sensor research and development performed by high energy physics researchers. This overlap is evident through the collaborations that have assembled under the umbrellas of the DOE Office of High Energy Physics' Quantum Information Science Enabled Discovery (QuantISED) program, DOE's National QIS Research Centers, and the National Science Foundations' Quantum Leap Challenge Institutes. Many of the techniques used by the QIS community are now being applied to challenges in high energy physics, such as particle theory quantum computation and methods for direct detection of dark matter. Likewise, device physics expertise from the high energy physics community is assisting the QIS community with elucidating some of the underlying processes that affect the performance of superconducting qubits. It is anticipated these research and development overlaps will continue in the coming decade, and some fraction of those efforts may take place or be demonstrated in underground facilities.

\subsection{Outlook}
As quantum computing systems advance in computational power and capability, radioactivity is likely to play an ever-increasing role in their performance.  Underground facilities, coupled with decades of experience of running sensitive, low-background experiments underground, offer a unique space to study how such systems behave in radiation-quiet environments.  These laboratories can also be a catalyst for further collaboration between the fields of particle physics, nuclear physics and QIS.  Investment in this line of research is strongly encouraged.

\section{Accelerator-based nuclear astrophysics}
Writer: Daniel Robertson (Notre Dame)

In the energy range of relevance to stellar burning the cross-sections of reactions of interest are extremely small. The rate of reaction drops very rapidly while moving lower into this energy range, reducing the measurable event rate below background. The nature of accelerator-based nuclear astrophysics experiments requires the active generation of radioactive decays through the forced interaction of nuclei of interest, this is in-order to detect the low-intensity signatures of thermonuclear reaction channels of significance in stellar burning environments. This process appears somewhat counter intuitive to being performed underground, where most other experiments relocate to avoid such event generation. The requirements to perform these interactions of interest however, show considerable overlap with numerous other fields moving underground for the same reason of background suppression.

\subsection{Science goals}
At the epicenter of the field of nuclear astrophysics is the drive to understand the synthesis of the elements in stellar environments~\cite{Burbidge_1957}. Understanding the generation of elemental abundances requires wide ranging information on stellar environments, particle interactions and energy generation. The current galactic elemental abundance is the result of numerous reaction paths all building on one another to create energy and elemental production. 

Initial production processes see the burning of hydrogen through either proton-proton interactions or through the CNO-cycle~\cite{Iliadis_2007} dependent on the mass of the star involved and available material present. Significant reactions of interest associated with the CNO cycle are the $^{12}$C(p,$\gamma$)$^{13}$N and $^{14}$N(p,$\gamma$)$^{15}$O, both of which are actively under investigation both above and below ground. The $^{14}$N(p,$\gamma$) reaction is of further significance through the first ddetection of $^{15}$O neutrinos by Borexino~\cite{Borexino_2020}. Continued burning passes through a helium phase where the triple-alpha process tales place, this key reaction of $^{12}$C($\alpha$,$\gamma$)$^{16}$O determines the ratio of carbon oxygen in our Universe and is considered the ''holy grail'' of nuclear astrophysics~\cite{deBoer_2017}. The complexity of this measurement and its obvious importance make it a prime candidate for underground measurements. 

Processes for the production of elements above the iron peak can be considered as two groups Group one for the production of elements far from stability and associated with multi-generational stars or explosive environments, \textit{r-process}~\cite{Cowan_2021}, \textit{p-process}~\cite{Arnould_2003}, \textit{rp-process}~\cite{Woosley_2004} and \textit{n-process}~\cite{Blake_1976}. Group two for the production of elements in non-explosive burning scenarios, \textit{i}-process~\cite{Aoki_2007} and s-process~\cite{Kappeler_2011}. Of specific interest for underground nuclear astrophysics is the s-process, where the slow capture of neutrons onto a seed nucleus creates almost half of all nuclei above mass 56 and along the valley of stability. Understanding and quantifying the source of neutrons for this process is of great interest to the community, the two main reactions thought to feed the neutron abundance are $^{13}$C($\alpha$,n)$^{16}$O and $^{22}$Ne($\alpha$,n)$^{25}$Mg. Significant efforts are underway to measure these reactions through direct and indirect methods, capitalizing on above ground and underground capabilities. Competing reactions which would reduce the abundance of $^{13}$C and $^{22}$Ne available for neutron production and those reacting with neutrons before s-process nuclei, known as neutron poisons, are also of strong importance. 

Recent white papers outlining the future direction of nuclear astrophysics~\cite{Aliotta_2022, Schatz_2022} have highlighted the current and expected future importance of underground reaction measurements. It is expected that underground facilities will grow even further into a significant tool for the measurement of key reactions of interest, with strong support from the nuclear astrophysics community. 

\subsection{Current underground-based research}
Ongoing research underground is strongly supported by its connection to above ground facilities performing higher energy measurements. These same facilities are striving to push the range of their measurements to lower energy through higher beam intensities, active shielding adaptations and more sophisticated detection techniques. The gains made through these methods help to better inform the requirements needed when inevitably the measurement must move underground. Hybrid facilities are operating tens of meters underground such as the Felsenkeller~\cite{Grieger_2020}, where some background suppression from rock overburden can be augmented with passive and active shielding, but this can only push reaction measurements so far into the region of interest.

In the US, underground nuclear astrophysics is pioneered by the CASPAR (Compact Accelerator System for Performing Astrophysical Research) collaboration, operating the only US-based deep underground low energy accelerator facility. Located at the 4850 level of the Sanford Underground Research Facility (SURF), the system is centered on a 1 MV Van de Graff style JN accelerator and has been fully operational since 2018~\cite{Robertson_2016}.  

Current studies at CASPAR are focusing on multiple avenues of CNO cycle physics and s-process neutron seeds. Due to the recent multi-disciplinary interest~\cite{Borexino_2020} a CASPAR (p,$\gamma$) campaign has measured the $^{14}$N(p,$\gamma$)$^{15}$O reaction~\cite{Frentz_2022} over a large energy range, generating a significant overlap with measurements above ground and previous underground investigations. The main focus however is in alpha capture reactions, ($\alpha$,$\gamma$) and ($\alpha$,n) with enhanced detection techniques of both $\gamma$ and neutron radiation. The big questions surrounding s-process neutron production and competing reactions are an on-going drive for future campaigns with necessary improvements on the already compelling work out there, for example $^{18}$O($\alpha$,$\gamma$)$^{22}$Ne ~\cite{Dombas_2022} and $^{22}$Ne($\alpha$,$\gamma$)$^{26}$Mg~\cite{Shahina_2022}. 

Further work is needed to reach deeper into the burning regime of interest. Specifically the upgrade of current equipment to cover larger energy ranges with increased beam intensity to target, the inclusion of more sophisticated detector techniques as a mechanism to cancel out natural background contamination from both detector materials and ambient rock, and a renewed effort by the community in target material production and the reduction of background producing contaminates. A number of this R\&D projects are actively being pursued above ground with anticipated extensions to underground work in the near future.           

Underground nuclear astrophysics was first established in 1992 when the LUNA (Laboratory for Underground Nuclear Astrophysics) collaboration installed their first 50 kV accelerator and measured the $^{3}$He($^{3}$He,2p)$^{4}$He reaction at solar energies~\cite{Junker_1998}. The group currently operates a 400 kV Singletron accelerator~\cite{Formicola_2003} allowing for a wider energy range of measurements including the ability to push into the energy range of interest for the $^{13}$C($\alpha$,$\gamma$)$^{16}$O reaction for the first time~\cite{Ciani_2021}. Commissioning work is underway to expand their capabilities with the addition of a new 3.5 MV accelerator and new laboratory space.

Further international efforts in China are centered around the JUNA (Jinping Underground Nuclear Astrophysics) facility located in the China Jinping underground Laboratory (CJPL)~\cite{Liu_2016}. The JUNA facility came online in 2021 and uses a high intensity ECR ion source and 400 kV platform to combine the background suppression available from the large rock overburden, with increased beam intensity at the lower energies required. Recent work has led to exciting results for the CNO cycle reactions with a direct measurement of the important $^{19}$F(p,$\alpha$)$^{16}$O reaction~\cite{Zhang_2021}.   

\subsection{Underground usage and needs}

\begin{itemize}
    \item $\textbf{Location:}$ As with other underground experiments, nuclear astrophysics laboratories make the move underground to suppress the high levels of cosmic induced background. The three deep underground facilities currently operating all make use of a rock overburden greater than 3500 m.w.e, however below $\sim$ 2000 m.w.e, the further reduction of cosmic induced background must be balanced against ambient backgrounds generated via uranium and thorium decay chains naturally occurring in the surrounding rock. This trade off must be studied per specific location, but may result in different levels of underground labs not suitable for other experiments, being useful to nuclear astrophysics. Secondary considerations stem from the general nature of the active production of nuclear reactions to generate radioactive decays, whereas the levels of production are intrinsically small, hence the need for underground, the production of possible secondary reactions during testing and R\&D stages may be undesirable to neighbouring experiments. Such interactions may be negated by either increased shielding at the accelerator facility or the consideration of more remote locations for future lab spaces.
    
    \item  $\textbf{Space requirements:}$ Required footprints for these underground accelerator facilities must consider not only the equipment hall, but at minimum a separated control room in close proximity. A compact system can be considered in as little as 4000 - 6000 sq ft, if suitable workshop and preparation space is available at nearby locations. Minimum height clearances for some high voltage platforms can be above 10 - 14 ft, but enclosed acceleration systems are modest if utilities can be located clear of working areas. Changing room facilities are also required, these are necessary as a transition zone between outer mine areas and lab spaces but do not have to meet the higher standards of clean room levels.      
    
    \item $\textbf{Cleanliness levels \& environment:}$ Air quality and climate controls for these accelerator based experiments are reliably uniform across different variations of systems. The transport of beam and interaction of nuclei of interest is all performed in vacuum, with only the detection system exposed to the ambient environment. However once these systems are exposed to air for configuration changes or maintenance, maintaining a clean, dry and dust free environment is necessary. Basic level requirements are outlined here, for future open-air high voltage platform installations it is felt that stricter controls maybe required dependent on the stability of humidity and dust levels. Most requirements overlap with those outlined by other groups.   
       \begin{itemize}
           \item $\textbf{Cleanroom}$ classification levels are not required for lab spaces, normal ``office'' air quality levels are sufficient with a need for any concrete floors to be sealed to reduce dust. A strong reduction of mine dust is a necessity at all entry and exit points to the main experimental halls and maintenance areas. 
           \item $\textbf{Humidity}$ levels are required between 30-50\% to reduce moisture uptake of vacuum systems and minimize condensation on water cooled equipment.    
           \item $\textbf{Vibration}$ levels are important for acceleration systems and detector set-ups and spaces should be considered at levels $<$ 100 micrometer/sec rms.
           \item $\textbf{Radon}$ mitigation is required at detector stations. Current systems utilize localized radon reduction boxes with a flow of dry nitrogen as a purge gas. A lab wide radon reduction infrastructure may be considered if beneficial to multiple groups.  
       \end{itemize}
    
    \item $\textbf{Utilities:}$ Infrastructure requirements are inline with other groups outlined in this section. Cooling water, uninterrupted power feeds, clean power at specified locations, high capacity wired and wireless network connections and liquid nitrogen supplies are all necessities.    
    
    \item $\textbf{Support and access:}$ All accelerator based experiments are strongly interactive and have significant access requirements. Whereas systems can be left to operate with more passive controls in place, start-up, shutdown and target changes need a physical presence. Access availability during campaigns must be considered on a flexible basis as 24/7 availability is a strong requirement for such systems.
    
\end{itemize}

\subsection{Outlook}
The outlook for the field of underground nuclear astrophysics experiments is extremely encouraging. As international efforts and upgrades are coming online new measurements continue to confirm the importance of key reactions and associated pathways in stellar models. A key challenge in nuclear astrophysics has always been the need to move away from the use of extrapolations from higher energies into astrophysically relevant burning regimes, reducing the uncertainties and thus refining stellar burning models for comparison and evaluation of observational data. Continued discoveries and refined measurements from these underground accelerator labs are used to guide research both above and below ground and are a driving force behind understanding the energy generation, elemental production and eventual life-cycle of stellar objects.  

Strong multi-disciplinary interaction and collaborations with current underground efforts is a necessity moving forward in developing and strengthening the filed of U.S. underground nuclear astrophysics. As highlighted throughout this $\textit{Underground Facilities}$ section there are many areas where requirements and capabilities can overlap and strengthen each other.

\section{Experiments in fundamental symmetries}
Writer: Catalina Curceanu (LNF, INFN)

\subsection{Science goals}
Refined measurements of X- and gamma-rays in an underground laboratory, represent an unique opportunity to investigate space-time related symmetries and the main conundrum of quantum theory, i.e. the collapse of the wave function. These investigations are being performed by extremely sensitive searches of atomic and nuclear transitions violating the Pauli Exclusion Principle (PEP), and as such, the spin-statistics connection, and of the spontaneous radiation predicted by the dynamical collapse models which solve the so-called ``measurement problem'' in quantum mechanics by adding non-linear terms to the Schroedinger equation. Possible violations of the Pauli Exclusion Principle are embedded in algebraic models, such as the quon model constrained to the Messiah-Greenberg (MG) superselection rule [1-4], and in Non-Commutative Quantum Gravity models (NCQG), common to both String Theory and Loop Quantum Gravity [5-9]. On the other side, the collapse models, such as the Continuous Spontaneous Localization and the gravity-related collapse models, have as a specific signature the so-called spontaneous radiation, i.e. a radiation with specific spectral features, depending on the collapse models parameters and on the emitting material, continuously emitted by the charged particles making up the matter (i.e. electrons and protons). The measurement of such radiation is a trademark of the collapse models [10-12].

\subsection{Current underground-based research}
Presently, the violation of the PEP is being investigated by various underground experiments, either by a dedicated search of the PEP violating atomic transition, such as the VIP experiment at the LNGS-INFN, or in experiments searching for other types of physics, such as DAMA/LIBRA or BOREXINO [13], [14] which used their data to also constrain the PEP. The VIP experiment singles out by having performed a dedicated search of PEP violation [15-16] with a method which allows to constrain the probability of the PEP violation by taking into account the Messiah-Greenberg superselection rule. The searches for the spontaneous radiation from collapse models was pioneered in the LNGS-INFN underground laboratory by the VIP collaboration, where the simplest version of gravity-related collapse models was ruled out [12], and tight constraints on CLS models were also set [11]. Presently, other underground experiments, not directly dedicated to the search of spontaneous radiation, are using their data to also try to constrain collapse models [17].

\subsection{Underground usage and needs}
Presently, searches of PEP violations are being performed by the VIP-2 collaboration at the LNGS in a dedicated experiment by circulating current through a copper (silver in next generation) target and searching for atomic transitions prohibited by PEP. Other underground experiments, not dedicated to PEP violations studies, measuring X- and gamma-rays, analyze their data in the view of searches for PEP violating nuclear and atomic transitions. The VIP-2 collaboration is also continuing to measure the  spontaneous radiation predicted by the collapse models in dedicated measurements with High Purity Germanium detectors at the low-radioactivity facility at LNGS. 

One can investigate the PEP violation and spontaneous radiation also in experiments measuring radiation with a different primary goal (dark matter, neutrino physics) as a synergistic activity; these experiments can analyze their data also to search for signals related to PEP violation and/or collapse models. However, there is not possible for these experiments to search for PEP violation taking into account the Messiah-Greenberg superselection rule [3], for which one needs to circulate a current into the target, so requiring a dedicated setup. In this case, one needs to adapt the setup by implementing a dedicated current source and target, as well as an additional shielding structure.

In the future, the VIP Collaboration is preparing a new version of the apparatus, VIP-3, to perform a more sensitive search, in the coming 3-5 years in the dedicated box at LNGS. This will be paralleled with dedicated searches of spontaneous radiation. The experimental needs are limited, both in terms of occupied space (basically a box), time duration and financing. One can envisage clever synergetic collaboration with other ongoing or planned experiments measuring radiation with a different primary purpose, which can be integrated/adapted to also perform searches for PEP violation and spontaneous radiation in various underground laboratories around the world. 

\subsection{Outlook}
The research performed by the VIP collaboration in searches of possible violation of PEP and of signals from collapse models in an underground laboratory, triggered theoretical research in the field which, in turn, is producing more refined models to be tested in the future. On one side, the PEP violation emerging from NGQG models requires a refined search of PEP violation as function of energy (i.e. along the periodic table). Recent theoretical studies are investigating the connection between PEP and CPT/Lorentz possible violations [8], which will require additional refined studies of PEP violation in dedicated experiments, eventually performed in various underground laboratories embracing various continents in Northern and Southern Hemispheres.

For the collapse models, under the pressure of recent published limits, new models are being developed, which go beyond the simplest versions by embedding dissipative and non-Markovian effects. All these models have specific spectral features which need to be investigated in future experiments. There is a huge window of opportunities for experiments in underground laboratories which could discover or set extremely tight constraints on theories which are impacting on our understanding of Nature, such as the spin-statistics connection and quantum mechanics. It is an excellent time to perform searches for PEP violation and collapse models in parallel with searches for dark matter and neutrino physics. The synergies in these research fields will enhance searches of radiation related to PEP violation and collapse models, impacting on our basic concepts rooting the entire modern science.

\section{Gravitational wave detection}
Writer: Jan Harms (GSSI, L'Aquila)

\subsection{Science goals}
Gravitational waves (GWs) carry an enormous amount of information about their sources and propagate virtually without loss through the Universe, which makes them a unique astrophysical and cosmological probe. Observations of GWs from mergers of binary neutron stars allow us to study extreme states of matter, and through multi-messenger studies of these sources, we can understand the formation of heavy elements and the engines of high-energy EM transients. The emission of GWs from coalescing black-hole binaries provides a unique probe of spacetime in the strong-curvature regime. It opens a new window into quantum-gravity effects and for the detection of new particles like axions accumulating near black-hole horizons. New sources of GWs might become observable in the future like core-collapse supernovae or rotating neutron stars. While some of the GW science still lies hidden in the noise of current GW detectors, a new generation of detectors will be able to access the full range of physics with GWs with high potential for breakthrough observations and a revolution in our understanding of the Universe~\cite{UF5-GW-Kalogera}.

\subsection{Current underground-based research}
Up to now, all GW observations were done with the LIGO and Virgo detectors, which are located on the surface. Surface environments are noisy in terms of seismic, atmospheric, and electromagnetic phenomena, which pose a limit to the sensitivity of low-frequency GW detections~\cite{UF5-GW-Fiori}. The KAGRA detector is located in a quiet underground environment and has recently joined the global detector network. It is projected to eventually exceed the low-frequency sensitivity of surface detectors. To unlock the full potential of underground GW detection, specific low-frequency instrument designs are required as planned for the proposed Einstein Telescope in Europe~\cite{UF5-GW-ET}. Such designs foresee the implementation of quantum back-action evading techniques, cryogenic technologies for the test masses and their suspensions, and further reduction of environmental noise using a new generation of isolation systems and advanced noise cancellation. It is worth noting there is substantial benefit to correlated and complementary observation from multiple GW detectors---inclusively above and below ground---to verify detection, refine direction, and observe details of the signal.

\subsection{Underground usage and needs}
The KAGRA detector will remain the only underground GW detector facility for the next decade at least to operate together with the LIGO and Virgo detectors~\cite{UF5-GW-Akutsu}. It is providing crucial insight into the operation and maintenance of underground GW detectors. For the breakthrough science promised with next-generation facilities like Cosmic Explorer and the Einstein Telescope, it will be necessary to construct detectors with greatly increased lengths of the interferometer arms compared to current detectors. Cosmic Explorer is planned as surface detector, but having its ends located inside mountains is still under consideration to benefit from reduced environmental noise. The Einstein Telescope is the only currently planned next-generation underground facility. It is proposed with a 10\,km arm length in the shape of an equilateral triangle. In preparation of a site selection and to inform its final underground infrastructure and detector designs, extensive studies of underground environments are being carried out at its candidate sites, but also elsewhere in the world including the KAGRA site. It is of great importance to understand the impact of service plants on underground infrastructure noise. In this regard, other types of underground research facilities like the Sanford Underground Research Facility and the National Laboratories of Gran Sasso can provide key insights, and analyses of data from these facilities have begun.

\subsection{Outlook}

The quest to operate GW detectors underground is entirely connected to the goal of extending the observation band of terrestrial detectors from currently about 20\,Hz to a few Hz in the future. Gravitational-wave observations in the few Hz to 20\,Hz band will enable the detection of mergers with intermediate-mass black holes up to a few 1000~solar masses. They are thought to have played a key role in the formation of the Universe’s large-scale structure as seeds of supermassive black holes and in the dynamics of star clusters, but their population is poorly understood today. Similarly, it will be possible to observe black-hole binaries to much larger redshift and to identify a possible primordial black-hole population. Even the analyses of less-massive sources like neutron-star binaries will greatly profit from the low-frequency sensitivity, which makes it possible to observe these sources for several hours. The Einstein Telescope is predicted to detect a few 100,000~GW signals per year up to redshifts of 100 with signal-to-noise ratios improved by more than an order of magnitude compared to current detectors, which will lead to the most extensive cosmological study of the Universe so far~\cite{UF5-GW-Maggiore}.

\section{Geology and geophysics}
Writers: Derek Elsworth, William Roggenthen, Herb Wang

\subsection{Science goals}
Underground research laboratories (URLs) offer opportunities for research that either cannot be performed at the surface or that can be performed more effectively in a URL. Goals include physics-based understanding of controls on permeability, stress, temperature, and chemical and biological processes across a variety of time scales ranging from fractions of a second to years and through both passive- and active-experimentation at length-scales of the order of 5-50m. Such experiments accommodate the effects of structure, heterogeneity, and fracturing/faulting at all scales under environmental conditions (\textit{e.g.}, stress, temperature) and with coupled processes relevant to engineered and natural geophysical processes with the ability to monitor and observe processes directly \textit{in situ}.

\paragraph{Permeability} Permeability, both natural and induced, is critical in a wide variety of applications. For example, the permeability of fault zones cannot be studied easily in the laboratory, but fluid transmissivity of faults is important in such diverse problems as the sealing of reservoirs, waste disposal strategies, geothermal energy production, and subsurface fluids (\textit{e.g.}, carbon dioxide, natural gas, and hydrogen) storage. 

\paragraph{Stress} Stress is a fundamental parameter for understanding the behavior of rocks in the subsurface. Development of improved measurement techniques is an important goal, especially in the confirmation and improvement of modeling approaches for the prediction of time dependent stresses. 

\paragraph{Temperature, Stress, and Chemical Processes} Combined effects of temperature, stress, and chemical processes involves changes in the \textit{in situ} conditions that result in alteration of the behavior of the rock to the passage of fluids, changes in the chemical composition of the rock, and mechanical behavior that are important in nuclear waste disposal and enhanced geothermal energy reservoirs. 

\paragraph{Rock Mass Characterization} Geophysical techniques for rock mass characterization further the development of rock mass characterization using techniques such as various tomography methods, \textit{e.g.} seismic and electrical resistivity, to allow prediction of underground stability and longevity of excavations.

\paragraph{Faulting, Fracturing, and Seismicity} Rock failure occurs over many spatial and temporal scales both from natural and engineered causes. Microearthquakes (MEQs) are a ubiquitous feature in the subsurface, accompanying underground excavation, construction, fluid injection or extraction, and other active experimentation. But large, induced earthquakes can also occur along with MEQs. Inducing and observing benign MEQs in URLs present useful analogs to understand modes of initiation and progress in natural and triggered events.

\subsection{Current underground-based research}
Current research is diverse and includes areas such as carbon sequestration, geothermal development, nuclear waste disposal issues, induced seismicity, and advances in underground excavation. The large physics-related URLs typically have aspects of geoscience and geoengineering involving research in rock mechanics that are necessary to ensure rock stability of the excavations. URLs such as the Sanford Underground Research Facility (SURF) are well-positioned to support the wider interests of the geosciences and geoengineering community and host geophysics, ground water, and geothermal experiments, as well. In addition to opportunities presented by URLs developed for physics investigations, research in many URLs is aimed toward work on issues relevant to nuclear waste disposal, such as that conducted at the Grimsel Test Site and Mont Terri Project in Switzerland and the WIPP Site in New Mexico, USA. The status of active, non-active, and planned URLs with this focus by Tynan and others (``\textit{A Global Survey of Deep Underground Facilities; Examples of Geotechnical and Engineering Capabilities, Achievements, Challenges\ldots}'', 2018) shows that thirteen of these types of URLs are currently active with more in the planning stage. In many instances, these facilities also provide platforms for needed advances in fluid flow through rocks and underground rock stability although none appear to have the capability of supporting experiments with extensive excavations at depths greater than 1~km.

\subsection{Underground usage and needs}
Typical installations consist of drilling of boreholes with associated instrumentation. At any one URL, an estimate of three such installations for a total of $\sim$750~m$^2$ of habitable space would be required. Any facility would desire borehole access to the adjacent geologic host---with characteristics selected that are dependent on the investigation---\textit{viz.} varied stresses, pressures, and temperatures (typically controlled by depth), specifics of rock structure, heterogeneity and fracturing and specifics of the stress field (obliquity \& orientation). These features are all key in aligning the characteristics of the URL experiment with those of the natural or engineered geological prototype. In addition to providing site access, necessary facilities include access to local workshops, qualified onsite technical personnel and management to support science, electrical power, water, compressed air, and ethernet access. 

\subsection{Outlook}
The outlook for mid-scale and highly constrained in-situ experiments in URLs is strong and are especially pertinent to societal needs such as those associated with energy and the environment. The Dept. of Energy has shown strong support for research in geothermal energy at SURF and carbon sequestration is of interest to both DOE and NSF. The initiatives are consistent with the NSF grand challenges, such as Growing Convergence Research and Mid-scale Research Infrastructure. The applied nature of the research is of interest to a range of industries, as well. 








\end{document}